\documentclass[conference]{IEEEtran}
\IEEEoverridecommandlockouts
\usepackage{graphicx}
\usepackage{amsmath}
\usepackage{amssymb}
\usepackage{booktabs}
\usepackage{multirow}
\usepackage{xcolor}
\usepackage{cite}
\usepackage{url}
\usepackage[hidelinks]{hyperref}

\IfFileExists{tables/results_macros.tex}{

\providecommand{\numClasses}{14}
\providecommand{\numFamilies}{3}
\newcommand{\totalImages}{10{,}316}

\newcommand{\flatBest}{ConvNeXt-Tiny}
\newcommand{\flatBestTopOne}{85.5}

\newcommand{\hierTopOne}{85.1}

\newcommand{\lOneAcc}{94.1}
\newcommand{\shortcutMaskDelta}{-2.9}
\newcommand{\shortcutCropDelta}{-0.7}

}{}
\IfFileExists{tables/cost_macros.tex}{\newcommand{\costK}{5}
\newcommand{\costReview}{0.5}
\newcommand{\flatCrossRate}{4.9}

\newcommand{\costOptReview}{21}
\newcommand{\costOptCost}{0.188}
\newcommand{\acceptAllCost}{0.342}
\newcommand{\eceBest}{0.037}
\newcommand{\aurcBest}{0.059}
\newcommand{\routingSavingPct}{45}
\newcommand{\hierCrossRate}{5.1}

}{}
\IfFileExists{tables/aux_macros.tex}{\newcommand{\clipTopOne}{13.5}
\newcommand{\ocrFinding}{on 200 test swatches OCR recognised text on 50\% but a construction keyword on only 0\%, so the text stream cannot disambiguate the visual prediction}
}{}
\providecommand{\numClasses}{14}
\providecommand{\numFamilies}{three}
\providecommand{\totalImages}{10{,}316}

\providecommand{\flatBest}{ConvNeXt-Tiny}
\providecommand{\flatBestTopOne}{XX.X}

\providecommand{\hierTopOne}{XX.X}

\providecommand{\lOneAcc}{XX.X}

\providecommand{\shortcutCropDelta}{XX}
\providecommand{\shortcutMaskDelta}{XX}
\providecommand{\eceBest}{0.0XX}
\providecommand{\aurcBest}{0.0XX}
\providecommand{\costK}{5}
\providecommand{\costReview}{0.5}
\providecommand{\flatCrossRate}{XX.X}
\providecommand{\hierCrossRate}{XX.X}

\providecommand{\costOptReview}{XX}
\providecommand{\clipTopOne}{XX.X}
\providecommand{\ocrFinding}{OCR recognises text on a minority of swatches, and essentially none of it names a construction, so the text stream cannot disambiguate the visual prediction.}

\begin{document}

\title{From Benchmark to Deployment: Shift-Robust\\
Fabric Recognition for Industrial Textile Onboarding}

\author{\IEEEauthorblockN{Haochen Li\textsuperscript{1,2}, Chenwei Wang\textsuperscript{1}, Felicity S.C. Tang\textsuperscript{1}, Misbah Iqbal\textsuperscript{1}, Carman K. M. Lee\textsuperscript{1,3}, Elif Ozden Yenigun\textsuperscript{1,2,*}}
\IEEEauthorblockA{\textsuperscript{1}Laboratory for Artificial Intelligence in Design, Hong Kong, China\\
\textsuperscript{2}School of Design, Royal College of Art, London, UK\\
\textsuperscript{3}Dept. of Industrial and Systems Engineering, The Hong Kong Polytechnic University\\
\textsuperscript{*}Corresponding author: elif.ozden-yenigun@rca.ac.uk}}

\maketitle
\thispagestyle{empty}
\pagestyle{empty}

\begin{abstract}
Automatically recognising a fabric's construction (jersey, twill, satin) is a bottleneck in
textile sourcing, where incoming swatches are still typed by hand. Benchmark accuracy suggests
the problem is solved, yet rarely survives deployment. On the \numClasses{}-class FabricFlow
benchmark we expose three gaps that headline accuracy hides. First, a duplication audit reveals
train/test leakage that inflates accuracy; we rebuild leakage-free splits that report the true
difficulty. Second, on the clean data the binding failure is acquisition-source shift between
catalogues, not the peripheral shortcuts one might fear: on an archive-exclusive hold-out,
standard training holds 58.0\% Top-1 at a calibration error of 0.158, while a simple,
architecture-agnostic central-texture recipe adds 13.5 Top-1 points and restores calibration.
Third,
because confusing one fabric family for another is costlier than a within-family slip, we
optimise a taxonomic-severity cost: a confidence-gated routing policy auto-types confident
swatches and refers only the uncertain minority to a human, sharply cutting onboarding cost.
Throughout we report honest negatives: hierarchical classification, OCR fusion and zero-shot
vision--language models all fail to help, yielding a concrete, calibrated, cost-aware recipe for
deployable textile onboarding.
\end{abstract}


\section{Introduction}
\label{sec:intro}

The textile and apparel industry is digitising, with growing demands for supply-chain
transparency, sustainability reporting, and automated quality control~\cite{industry_ai,industry5_breque,bertola2018fashion40}.
An early, consequential step is \emph{material onboarding}: assigning each incoming fabric
swatch a construction label (plain weave, jersey, twill, \dots)~\cite{textileterms2002} that feeds procurement, pricing,
quality assurance, and traceability. Today this is overwhelmingly manual: slow, inconsistent
between operators, and a bottleneck as material libraries scale to thousands of SKUs.

Sourcing platforms already photograph every swatch, so computer vision is a natural lever: fabric
texture and material appearance are learnable from ordinary photographs~\cite{tan2024densenet_fabric,material_recog}. But
a leaderboard accuracy is not a deployable system. Moving a recogniser into a textile-onboarding
line surfaces three requirements that a single accuracy number hides, and this paper addresses
all three on the FabricFlow benchmark:
\begin{enumerate}
\item \textbf{Trust the data.} Reported accuracy must reflect fabric texture, not dataset
leakage or peripheral artifacts.
\item \textbf{Survive distribution shift.} The model must hold up when the acquisition
conditions change between catalogues, not only on an i.i.d.\ test split~\cite{koh2021wilds}.
\item \textbf{Optimise the right objective.} In onboarding, errors are not equally costly:
mistyping a knit as a woven (cross-family) routes a swatch into the wrong material workflow,
whereas a within-family slip is cheap; the system must also know when to defer to a human.
\end{enumerate}

Our contributions follow this arc.
\textbf{(1) A validity-controlled benchmark.} Because dataset bias and near-duplication routinely
inflate reported accuracy~\cite{torralba2011bias}, a perceptual-hash plus byte-level leakage audit
uncovers $\sim$29\% duplicated train/test images (inflating a true $\sim$86\% to 91\%); we
rebuild leakage-free cluster-level splits, and a de-confounded crop$\times$mask factorial shows
the curated swatches carry few exploitable peripheral shortcuts.
\textbf{(2) Source-shift robustness.} On the clean splits the binding failure is
acquisition-source shift rather than shortcuts: on an archive-exclusive hold-out, standard
training reaches 58.0\% Top-1 at a calibration error of 0.158, and a simple,
architecture-agnostic \emph{central-texture} recipe adds 13.5 Top-1 points (95\% bootstrap CI
$[+11.9,+15.2]$) and cuts calibration error to 0.028.
\textbf{(3) Cost-aware deployment.} Under a taxonomic-severity cost model, a cost-optimal
confidence-gated routing policy cuts expected onboarding cost by \routingSavingPct\%; we further
show, counter-intuitively, that a taxonomy-aware hierarchy gives \emph{no} cost advantage.
\textbf{(4) Honest negatives} (OCR fusion, zero-shot CLIP, and the hierarchy itself) that tell
practitioners what \emph{not} to build. Novelty is claimed at the level of this deployment-oriented integration and
its findings, not a new neural architecture.

\section{Related Work}
\label{sec:related}

\paragraph{Fabric vision in industry}
Most textile deep-learning targets defect detection and surface
inspection~\cite{defect_survey,kahraman2023fabric_review,liu2020multistagegan} or single-source weave/pattern
classification~\cite{hussain2020woven,tan2024densenet_fabric}, within the Industry~4.0/5.0 move toward
machine-learning-based quality control~\cite{industry_ai,industry5_breque,bertolini2021ml_industrial,villalba2019industry_cv}.
Class definitions follow standard textile references~\cite{spencer2001knitting,textileterms2002,iso8388}. Deployable
construction-type recognition that is leakage-audited, shift-robust, and cost-aware is
under-addressed; that is the gap we target.

\paragraph{Dataset bias, leakage, and distribution shift}
Dataset bias and train/test leakage inflate in-distribution accuracy and mask poor
transfer~\cite{torralba2011bias,koh2021wilds}; near-duplicate purging is the established
remedy~\cite{barz2020duplicates}, so we audit and deduplicate before measuring.
Robustness under shift is addressed by domain generalisation, e.g.\
MixStyle~\cite{zhou2021mixstyle}, and by test-time adaptation, e.g.\ entropy minimisation
(TENT)~\cite{wang2021tent}; MixStyle assumes several labelled source domains and TENT assumes
online adaptation on target batches, so we keep the mitigation at inference time on a single
training domain.

\paragraph{Acquisition-conditioned recognition beyond textiles}
Fabric onboarding is not the only setting in which appearance is a function of how the image was
captured, and the responses developed in other acquisition-limited domains map onto the three
requirements above. Where the sensing geometry dominates object appearance and labelled data are
scarce, recognition has been hardened by isolating the class-crucial evidence from acquisition
nuisance~\cite{torralba2011bias,wang2023crucial}, by causal intervention that removes the
confounding of acquisition conditions from the class
evidence~\cite{geirhos2020shortcut,wang2024unveiling,wang2025limited}, by generative augmentation
when the capture conditions cannot simply be
resampled~\cite{jeong2023kimageexpansion,wang2022sar}, and by auxiliary segmentation that tells a
recogniser where the object is rather than what surrounds
it~\cite{jocher2023yolo,wang2020deep,wang2021deep,wang2022semi}. Aggregating several views of the
same object before deciding is a recurring device in that
literature~\cite{shanmugam2021tta,wang2020multi,wang2021multiview} and is the antecedent of the
multi-crop test-time averaging of Section~\ref{sec:method}. Structuring the decision is another
shared response: coarse-to-fine label hierarchies and feature refiners that reuse the coarse
stage~\cite{zheng2020progressive,wang2022recognition,wang2023sar1,shang2023hdss,chenwei_taha_inprep},
attention pooled over several scales~\cite{rodriguez2020attention,wang2023sar}, and hybrid
convolutional--transformer designs that keep global context under few-shot
budgets~\cite{dosovitskiy2021vit,wang2022global}; the accumulated practice for the
acquisition-limited case is surveyed at book length in~\cite{wang2026synthetic}. Deciding when
\emph{not} to decide is pursued there as well, through open-set recognition and through detectors
for inputs that fall outside the training distribution, whether corrupted, multimodal, or scored on
device~\cite{hendrycks2019benchmarking,wang2023entropy,lili2025dpu,lili2025secure}. Deployed
industrial sensing systems --- inspection-activity
recognition~\cite{multimodal_inspection,yin2025spatio,yin2026ciuav}, infrastructure condition
monitoring~\cite{he2021textile_drl,luo2022evaluating} and in-vehicle
perception~\cite{villalba2019industry_cv,wang2019parking} --- make the same point from the
deployment side: the binding constraint is rarely the classifier but the operating conditions and
the cost of being wrong. Domain-specific imaging pipelines in the life sciences, where restoring or
standardising the acquisition precedes any
analysis~\cite{zhuang2021transfer_survey,li2025volume,guan2025cell}, follow the same ordering of
priorities as our leakage audit and central-texture recipe. The validity concern behind that audit
recurs wherever labels are assembled at scale: annotation bias has to be identified before it is
trained on~\cite{torralba2011bias,li2023biased}, and prototype-based and domain-invariant
formulations are the standard response when the label space is structured and the capture
conditions vary~\cite{zhou2021mixstyle,lili2024panoptic,lili2024domain}.

\paragraph{Recognition methods}
Distinguishing fine-grained fabric classes that share texture is a fine-grained recognition
problem~\cite{wei2022finegrained}, addressed with attention/relation
modules~\cite{liu2022crosspart,zhao2017diversified,rodriguez2020attention,bera2022srgnn,chang2020mutualchannel}; classical texture descriptors remain
informative~\cite{texture_classic,liu2016mrelbp,guo2016sslbp,zhang2018texture}. Coarse-to-fine hierarchical
classifiers~\cite{hierarchical_survey,gou2024hierarchical} motivate our knit/woven cascade, kept standard for
reproducibility. We compare convolutional, transformer, and lightweight
backbones~\cite{he2016resnet,liu2022convnext,liu2021swin,touvron2021deit,howard2019mobilenetv3};
because architecture family itself modulates robustness under
shift~\cite{shukla2025robustnessbench}, this cross-family sweep is part of the shift protocol
rather than a leaderboard.

\paragraph{Cost, calibration, and selective prediction}
Cost-sensitive learning weights errors by their consequences~\cite{elkan2001cost}, and deep
formulations learn class-dependent costs jointly with the
representation~\cite{khan2018costsensitive}; we
instantiate this with a taxonomy-derived severity matrix. Deployment reliability rests on
calibration~\cite{guo2017calibration,minderer2021calibration}, whose degradation under dataset
shift is well documented~\cite{ovadia2019trust}, and selective prediction / classification with a reject
option~\cite{geifman2017selective,geifman2019selectivenet}, which we combine into a cost-optimal
routing policy. An OCR/text stream~\cite{shi2017endtoend} and a zero-shot vision--language
model~\cite{radford2021clip} are evaluated as auxiliary references; both are unhelpful here, even
though image--text fusion helps fine-grained discrimination in other
domains~\cite{xu2024finegrained,mai2023mib,zhang2023vpgtrans}. A companion system on the same benchmark pursues that
direction further, arbitrating several vision--language models under an explicit cost
budget~\cite{chenwei_vlmarbiter_inprep}.

\section{Benchmark and Validity}
\label{sec:dataset}

\paragraph{Taxonomy and composition}
FabricFlow comprises \totalImages{} cleanly-decoding commercial fabric-swatch photographs over
\numClasses{} construction classes in \numFamilies{} families: KNIT (Jersey, Rib Knit, French
Terry, Interlock, Tricot, Knit Jacquard), WOVEN (Plain Weave, Twill, Satin, Corduroy, Ribbed
Poplin, Leno Gauze, Woven Jacquard), and OTHERS (Mesh/Lace). The label space follows textile
construction, not surface appearance~\cite{iso8388,textileterms2002}, so the two jacquards stay distinct across families.
Per-class counts are mildly imbalanced, so we report macro-averaged metrics and class-stratified
splits.

\paragraph{Leakage audit}
Before trusting any number we audit for leakage. A perceptual hash flags many test images as
near-duplicates of training images, following standard near-duplicate purging
practice~\cite{barz2020duplicates}; we confirm each flag with byte-level MD5 and pixel
correlation. The duplication is real and severe: $\sim$29\% of test images are byte-identical to
a training image because the nominal acquisition-source folders are partly copies of one
another, which inflates a true $\sim$86\% random-split accuracy to a reported 91\%. We therefore
cluster images by content and rebuild \emph{cluster-level} leakage-free splits in which a content
cluster never straddles folds: a leakage-free \emph{random} split (ten classes retain enough
distinct content for clean folds), and a leakage-free \emph{source} split whose test set holds
content appearing only in archival acquisition batches (the seven archive-exclusive classes). All
numbers below use these deduplicated splits; the leakage-inflated figures appear only to quantify
the inflation.

\paragraph{Shortcut control}
Commercial swatches can carry peripheral cues (labels, rulers, background) that a network can
latch onto as shortcuts~\cite{geirhos2020shortcut}. We treat preprocessing
as an experimental variable with two orthogonal controls, a centre \emph{crop} and an edge
\emph{mask}, and run a constant-augmentation crop$\times$mask factorial (Table~\ref{tab:shortcut})
so the main effects isolate each control. The result is informatively negative (crop
\shortcutCropDelta{}~pp, mask \shortcutMaskDelta{}~pp): the curated swatches carry few exploitable
peripheral shortcuts, so reported accuracy reflects texture rather than artifacts, and the
source-shift result below is therefore \emph{not} a shortcut artifact.
\begin{table}[t]
\centering
\footnotesize
\caption{De-confounded crop$\times$mask factorial (ConvNeXt-Tiny, constant augmentation). Main effects isolate each control from augmentation strength.}
\label{tab:shortcut}
\begin{tabular}{lrrr}
\toprule
Condition & Top-1 & Top-2 & macro-F1 \\
\midrule
No crop, no mask & 84.41 & 93.02 & 84.23 \\
Crop, no mask & 83.06 & 94.38 & 82.73 \\
No crop, mask & 80.93 & 91.60 & 80.51 \\
Crop, mask & 80.78 & 90.53 & 80.50 \\
\midrule
\multicolumn{4}{l}{Crop main effect: -0.75 pp\quad Mask main effect: -2.88 pp} \\
\bottomrule
\end{tabular}
\end{table}

\section{Method}
\label{sec:method}

\paragraph{Recogniser}
The base recogniser is a single ImageNet-pretrained backbone~\cite{zhuang2021transfer_survey} with a softmax classification head,
fine-tuned on the clean training split. We also evaluate a taxonomy-aware cascade as a candidate
for both accuracy and cost: an L1 router predicting the family (KNIT/WOVEN/OTHERS) feeds L2
within-family specialists, assembled by marginalising
$P(c\mid x)=P(\mathrm{fam}(c)\mid x)\,P(c\mid\mathrm{fam}(c),x)$.
Section~\ref{sec:results} shows it earns neither.

\paragraph{Central-texture inference for source-shift robustness}
The source shift we diagnose (Section~\ref{sec:results}) is driven by peripheral acquisition
artifacts, measuring rulers and altered framing, that the construction class does not depend on.
We therefore make inference read the central texture and ignore the periphery, with two
composable, architecture-agnostic ingredients: (i) \emph{strong scale and framing augmentation}
during fine-tuning, so the model is invariant to acquisition framing; and (ii) \emph{multi-crop
test-time averaging}~\cite{shanmugam2021tta}, which averages the softmax over three central crops covering 80\%, 65\% and
50\% of the frame. Neither requires source labels or target data.

\paragraph{Taxonomic-severity cost model}
For onboarding, errors are not equally costly~\cite{elkan2001cost}. We assign each (true $y$, predicted $\hat y$) pair
\begin{equation}
C(y,\hat y)=\begin{cases}
0 & \hat y=y,\\
1 & \hat y\neq y,\ \mathrm{fam}(\hat y)=\mathrm{fam}(y),\\
K & \mathrm{fam}(\hat y)\neq\mathrm{fam}(y),
\end{cases}
\label{eq:cost}
\end{equation}
with default $K{=}\costK$: a cross-family error costs $K\times$ a within-family one. A model's
\emph{expected cost} is $\bar C=\tfrac{1}{N}\sum_i C(y_i,\hat y_i)$, a deployment-relevant score
that penalises the operationally severe mistakes.

\begin{figure}[t]\centering
\includegraphics[width=0.82\columnwidth]{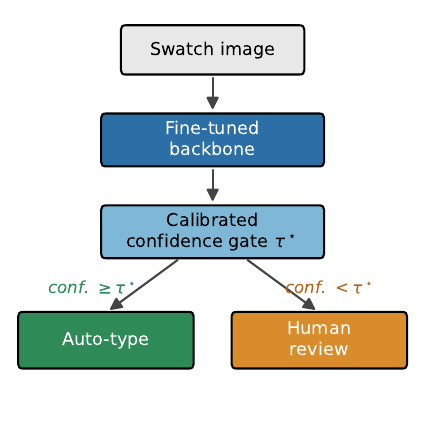}
\caption{Deployment pipeline. A fine-tuned backbone (with central-texture inference under shift)
feeds a calibrated confidence gate whose threshold $\tau^\star$ minimises expected onboarding
cost; confident swatches are auto-typed, the uncertain high-cost minority is routed to review.}
\label{fig:pipeline}
\end{figure}

\paragraph{Cost-optimal confidence-gated routing}
Auto-accepting a swatch incurs its realised cost; routing it to a human costs
$C_{\text{review}}{=}\costReview$. With $s_i$ the model's confidence (max softmax), accepting all
$s_i\ge\tau$ and routing the rest, the cost-optimal threshold is
\begin{equation}
\tau^\star=\arg\min_{\tau}\ \frac{1}{N}\Big[\!\!\sum_{i:\,s_i\ge\tau}\!\! C(y_i,\hat y_i)
\;+\; C_{\text{review}}\,\big|\{i: s_i<\tau\}\big|\Big],
\label{eq:route}
\end{equation}
trading residual misclassification cost against review effort (Fig.~\ref{fig:pipeline}). Backbones
are fine-tuned with AdamW~\cite{loshchilov2019adamw}, cosine decay, AMP, label
smoothing~\cite{szegedy2016labelsmooth,zhang2021labelsmoothing}, and early stopping; evaluation uses
horizontal-flip TTA~\cite{shanmugam2021tta}.

\section{Experimental Protocol}
\label{sec:experiments}

All experiments use the leakage-free cluster-level splits of Section~\ref{sec:dataset}: a
\emph{random} split (class-stratified, ten classes with clean coverage) for in-distribution
accuracy and the cost analysis, and a \emph{source} hold-out (seven archive-exclusive classes)
for the distribution-shift study; source-shift numbers are means over three seeds (42, 1, 2), with
$95\%$ bootstrap confidence intervals on the difference from ERM. The flat baseline runs across five backbones spanning
convolutional, transformer, and lightweight families (ConvNeXt-Tiny~\cite{liu2022convnext},
Swin-Tiny~\cite{liu2021swin}, DeiT-Small~\cite{touvron2021deit}, ResNet-50~\cite{he2016resnet},
MobileNet-V3-Large~\cite{howard2019mobilenetv3}), a cross-family sweep motivated by the
architecture dependence of robustness under shift~\cite{shukla2025robustnessbench}; ConvNeXt-Tiny is the primary backbone for the
shift, hierarchy, and cost studies. Training uses AdamW ($10^{-4}$), cosine decay, AMP, label
smoothing, batch 48, $224^2$ images, early stopping, on a single RTX~3070. We report Top-1/Top-2,
macro-F1, ECE, AURC, coverage at 95\% selective accuracy and, central here, the cross-family error
rate and expected onboarding cost. A zero-shot CLIP~\cite{radford2021clip} reference and an OCR ablation
complete the protocol.

\section{Results}
\label{sec:results}

\subsection{In-distribution baseline}
On the leakage-free random split, \flatBest{} is the strongest of five backbones at
\flatBestTopOne\% Top-1 (Table~\ref{tab:flat}); it is well calibrated (ECE \eceBest{}, AURC
\aurcBest{}). The taxonomy-aware cascade matches it (L1 router \lOneAcc\%, assembled
\hierTopOne\%) but does not beat it: its only edge is an interpretable decomposition. We fix
ConvNeXt-Tiny as the backbone below.
\begin{table}[t]
\centering
\footnotesize
\caption{In-distribution baselines across five backbones (14-class leakage-free random split, test set). Top-1/Top-2/macro-F1 in \%; ECE lower is better. ConvNeXt-Tiny is the backbone used throughout.}
\label{tab:flat}
\begin{tabular}{lrrrr}
\toprule
Backbone & Top-1 & Top-2 & macro-F1 & ECE \\
\midrule
ConvNeXt-Tiny & 85.48 & 94.38 & 85.23 & 0.037 \\
Swin-Tiny & 84.63 & 93.67 & 84.34 & 0.039 \\
DeiT-Small & 82.06 & 91.39 & 81.87 & 0.052 \\
ResNet-50 & 82.99 & 92.24 & 82.51 & 0.034 \\
MobileNet-V3-L & 81.71 & 90.68 & 81.30 & 0.052 \\
\bottomrule
\end{tabular}
\end{table}

\subsection{Acquisition-source shift and its mitigation}
\begin{figure}[t]\centering
\includegraphics[width=\columnwidth]{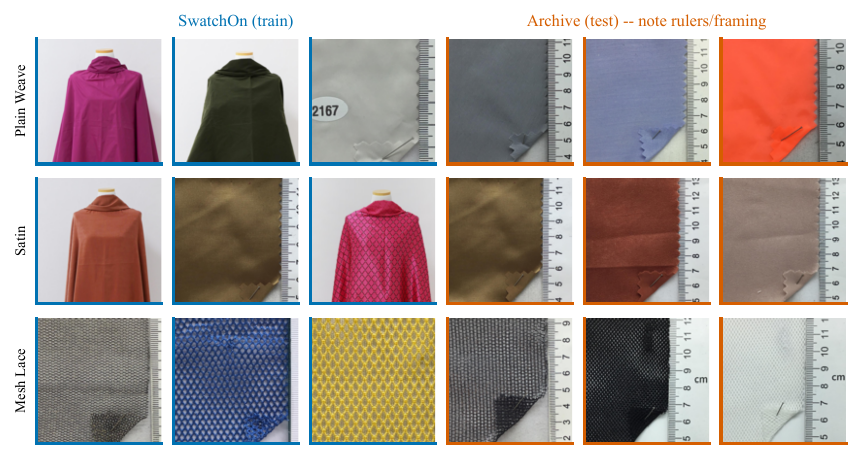}
\caption{The source shift is an acquisition-style shift. Training swatches (blue) are clean
flat-lays; archive-exclusive test swatches of the same classes (orange) carry peripheral
measuring rulers and altered framing the model never sees in training, which the central-texture
recipe is designed to ignore.}
\label{fig:shift}
\end{figure}

Holding out content that appears only in archival acquisition batches leaves ConvNeXt-Tiny at
$58.0\pm5.3\%$ Top-1 over the seven archive-exclusive classes, at ECE 0.158, with 1.6\% of
swatches auto-acceptable at 95\% selective accuracy (Table~\ref{tab:sourceshift}): under shift the
model is both inaccurate and unable to signal it. The mechanism is visible to the eye
(Fig.~\ref{fig:shift}): archival swatches carry peripheral measuring rulers and altered framing
the model never saw in training. The crop$\times$mask controls of Section~\ref{sec:dataset} move
accuracy by \shortcutCropDelta{} and \shortcutMaskDelta{}~pp, so on this benchmark the deployment
risk sits in the acquisition shift rather than in peripheral shortcuts.

The central-texture recipe attacks that mechanism without source labels or target data. Strong
scale and framing augmentation alone adds $5.3$ Top-1 points (95\% bootstrap CI $[+3.9,+6.7]$);
adding multi-crop test-time averaging takes the total to $+13.5$ points ($[+11.9,+15.2]$), for
$72.8\pm0.9\%$. Calibration moves with it, ECE $0.158\rightarrow0.028$, and the seed-to-seed
spread narrows from $\pm5.3$ to $\pm0.9$. For deployment the sharpest number is coverage: the
share of swatches clearing 95\% selective accuracy~\cite{geifman2017selective} rises from 1.6\% to 32.7\%, a twentyfold
increase in what an onboarding line can auto-type without review.
\begin{table}[t]\centering\footnotesize
\setlength{\tabcolsep}{4pt}
\caption{Acquisition-source shift on the leakage-free source split (2{,}928 test images from the
seven archive-exclusive classes; ConvNeXt-Tiny; means over three seeds). ECE: expected calibration
error. AURC: area under the risk--coverage curve, lower is better. Cov@95: share of swatches
auto-acceptable at 95\% selective accuracy. The Top-1 gain over ERM is $+5.3$ points for strong
augmentation (95\% bootstrap CI $[+3.9,+6.7]$) and $+13.5$ points once multi-crop averaging is
added ($[+11.9,+15.2]$); both intervals exclude zero.}
\label{tab:sourceshift}
\begin{tabular}{lrrrr}
\toprule
Inference recipe & Top-1 & ECE & AURC & Cov@95 \\
\midrule
ERM (standard training) & 58.0\,$\pm$\,5.3 & 0.158 & 0.265 & 1.6\% \\
Strong augmentation & 64.0\,$\pm$\,0.8 & 0.087 & 0.185 & 7.4\% \\
\quad + multi-crop TTA & \textbf{72.8\,$\pm$\,0.9} & \textbf{0.028} & \textbf{0.108} & \textbf{32.7\%} \\
\bottomrule
\end{tabular}
\end{table}

\subsection{Cost-aware deployment}
\begin{figure}[t]\centering
\includegraphics[width=0.82\columnwidth]{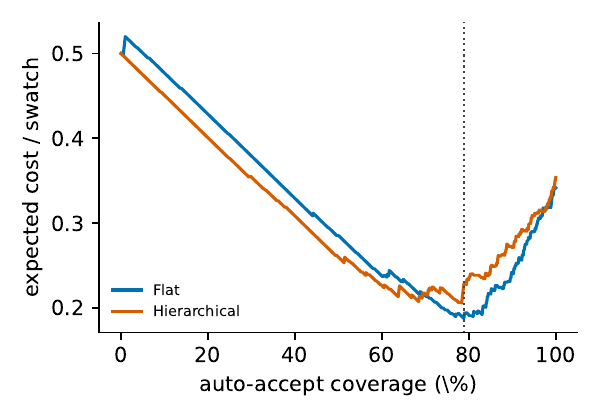}
\caption{Expected onboarding cost vs.\ auto-accept coverage; the minimum marks the cost-optimal
operating point. The hierarchy's curve lies on or above the flat model's (no cost advantage);
routing cuts cost far more than any architecture choice.}
\label{fig:cost}
\end{figure}

For onboarding, accuracy is the wrong objective. Under the taxonomic-severity cost model
($C_{\text{cross}}{=}\costK\times$ within-family, $C_{\text{review}}{=}\costReview$;
Table~\ref{tab:cost}) we first test whether the hierarchy, via its accurate family router,
commits fewer costly cross-family errors. \emph{It does not}: both models confuse families only
rarely (\flatCrossRate\% flat vs.\ \hierCrossRate\% hierarchy) and the hierarchy's expected cost
is marginally \emph{higher}, so the flat head is already family-robust and an explicit L1 stage is
redundant. The cost win comes from \emph{routing} (Fig.~\ref{fig:cost}): ranking by confidence and
choosing the coverage that minimises total expected cost auto-types the confident majority and
routes only $\sim$\costOptReview\% (the uncertain, high-cost minority) for review, cutting expected
cost from \acceptAllCost{} to \costOptCost{} per swatch, a \textbf{\routingSavingPct\% reduction},
robust across $K\in\{2,5,10\}$. The practical guidance inverts the intuitive one: deploy the
simpler flat model and invest in the cost-tuned gate, not the hierarchy.
\begin{table}[t]
\centering
\footnotesize
\caption{Cost-aware deployment (14-class leakage-free test). Cross-family errors cost 5$\times$ a within-family error; human review costs 0.5 per swatch. Both models rarely confuse families ($\sim$5\% cross-family error), so the taxonomy-aware hierarchy gives \emph{no} cost advantage over the flat model; the value comes instead from the cost-optimal confidence-gated routing policy (Fig.~\ref{fig:cost}).}
\label{tab:cost}
\begin{tabular}{lrrrr}
\toprule
Model & Top-1 & X-fam.\ (\%) & Cost & Review (\%) \\
\midrule
Flat (convnext-tiny) & 85.48 & 4.91 & 0.342 & 21.1 \\
Hierarchical (soft) & 85.12 & 5.12 & 0.354 & 21.6 \\
\bottomrule
\end{tabular}
\end{table}

\subsection{Honest negatives}
An OCR/text stream adds nothing: on the swatches, \ocrFinding{} A zero-shot CLIP classifier reaches
only \clipTopOne\%, far below the fine-tuned models, so domain-specific fine-tuning is not
optional~\cite{zhuang2021transfer_survey}.
Together with the hierarchy, these results give a clear ``what not to build'' list for a
fabric-onboarding line.

\section{Discussion}
\label{sec:discussion}

\paragraph{A deployment recipe, not a leaderboard entry}
The three pillars compose into a concrete procedure for a textile-onboarding line. First,
\emph{audit the benchmark}: leakage and shortcut controls establish that the headline accuracy
reflects fabric texture, so everything downstream rests on real signal. Second, \emph{harden
against the real shift}: the failure that binds is not the artifact one might fear (peripheral
shortcuts) but acquisition-source shift, and a cheap, architecture-agnostic central-texture recipe
adds 13.5 Top-1 points on the shifted fold and restores calibration, with no source labels or
target data required. Third, \emph{operate by cost}: a calibrated confidence gate tuned to a
taxonomic-severity cost auto-types the confident majority and escalates the uncertain, high-cost
minority, cutting expected onboarding cost by \routingSavingPct\% under a stated review budget.

\paragraph{What not to build}
Equally useful for practitioners is what our honest negatives rule out. A multi-stage taxonomy
hierarchy buys neither accuracy nor cost over a single flat head (the families are already
visually separable); an OCR/text stream cannot fire (swatch text rarely names a construction);
and a zero-shot vision--language model trails fine-tuning by a wide margin. The deployable system
is therefore deliberately simple: one fine-tuned backbone, central-texture inference, calibration,
and a cost-tuned gate.

\paragraph{Generality and scope}
The method (scoring and operating a classifier by a taxonomy-derived severity cost, hardening it
with central-texture inference, and routing the high-cost-uncertain tail to a human) transfers to
any onboarding or inspection task with operationally unequal errors and acquisition variation.
The benchmark is single-platform, so the shift studied here is a within-platform acquisition shift
between catalogues rather than a cross-supplier guarantee, and the shift results are measured
against ERM on the same archive-exclusive fold, which bounds what the recipe recovers rather than
the absolute cost of the shift. One backbone and three seeds carry that study; a multi-source
trial with a matched in-distribution control is the natural next step.

\section{Conclusion}
\label{sec:conclusion}

We presented a deployment-oriented study of fabric recognition for industrial textile onboarding
on the FabricFlow benchmark, following the path from benchmark to deployment. A leakage audit
removes $\sim$29\% duplicated images and a shortcut-control factorial establishes that accuracy
reflects texture. On the resulting leakage-free splits, acquisition-source shift rather than peripheral
shortcuts is what limits deployable accuracy: standard training holds only 58.0\% Top-1 on an
archive-exclusive fold, and a simple central-texture recipe lifts it to 72.8\% ($+13.5$\,pt) while
cutting calibration error from 0.158 to 0.028. Reframing the objective around industrial cost, a
cost-optimal confidence-gated routing policy cuts expected onboarding cost by \routingSavingPct\%,
whereas a taxonomy-aware hierarchy gives no cost advantage. With OCR and zero-shot CLIP reported
as honest negatives, the takeaway is a concrete, calibrated, shift-robust, cost-aware recipe, and
a clear ``what not to build'' list, for deployable textile onboarding. Future work includes a multi-source
deployment trial and a richer cost matrix.

\section*{Acknowledgment}
This research is funded by the Laboratory for Artificial Intelligence in Design (Project Code: RP2-2) under the InnoHK Research Clusters, Hong Kong Special Administrative Region Government. The authors thank Buoyuan Tuo for collecting the SwatchOn fabric swatch dataset used throughout this study.

\bibliographystyle{IEEEtran}
\bibliography{refs}
\end{document}